\documentclass[twocolumn,showpacs]{revtex4}
\usepackage[xdvi]{graphicx}%
\usepackage{dcolumn}
\usepackage{amsmath}
\makeatletter

\def\btt#1{\texttt{\@backslashchar#1}}%
\DeclareRobustCommand\bblash{\btt{\@backslashchar}}%
\makeatother

\newcommand{\bra}{\left\langle}
\newcommand{\ket}{\right\rangle}

\newcommand{\pder}[2]{\frac{\partial #1}{\partial  #2}}

\begin{document}

\title{Thermodynamic relations in a driven lattice gas:
numerical experiments}
\author{Kumiko Hayashi and Shin-ichi Sasa}
\affiliation
{Department of Pure and Applied Sciences,
University of Tokyo, Komaba, Tokyo 153-8902, Japan}

\date{\today}

\begin{abstract}
We explore thermodynamic relations in non-equilibrium steady states with 
numerical experiments on a driven lattice gas.  After operationally 
defining the pressure and chemical potential in the driven 
lattice gas, 
we confirm numerically the validity of the integrability condition 
(the Maxwell relation) for the two quantities whose values differ
from those for an equilibrium system. 
This implies that a free energy function 
can be constructed for the non-equilibrium steady state that we consider.
We also investigate 
a fluctuation relation associated with this free energy function. Our result 
suggests that the compressibility can be expressed in terms of
density fluctuations even in non-equilibrium steady states.
\end{abstract}

\maketitle

A rich variety of non-equilibrium phenomena have been successfully described 
by phenomenological evolution equations. However, the microscopic 
foundation of such equations has not yet been established, 
except for systems near equilibrium.
Even for non-equilibrium steady states (NESS) realized in simple systems,
such as those involving only heat conduction and shear flow, 
appropriate statistical 
measures  of microscopic configurations are not known. 
Recalling  that equilibrium statistical mechanics was constructed with 
the aid of thermodynamics, we expect that checking the validity  of 
thermodynamic relations in NESS is an important  step 
in constructing a theory of non-equilibrium statistical mechanics.

Non-equilibrium lattice gases  are simple mathematical models 
which have been useful in the elucidation of 
universal properties of NESS \cite{LG}. 
Topics studied with such models include
nonequilibrium phase transitions \cite{KL}, 
long-range spatial correlations \cite{GL}, fluctuation theorems \cite{LS}, 
and non-local large deviation functionals \cite{DLS}, as well as 
mathematical foundations of nonequilibrium  statistical mechanics 
\cite{Spohn}. It is thus expected that the non-equilibrium lattice gases
provide good models for the exploration of thermodynamic relations.

There have been some proposals of an extended  thermodynamic framework
applicable to NESS 
\cite{Oono,ST}.  In one such study, Sasa and Tasaki 
start from operational definitions of  the pressure $p$ and 
chemical potential $\mu$, and they derive from these a quantitative 
relation which can be tested experimentally \cite{ST}. 
Because the Maxwell relation for $p$ and $\mu$ plays an essential role in 
the derivation of this relation, we are led to study the same Maxwell 
relation in the case of a driven lattice gas.  

In this Letter, we present numerical results that confirm the validity
of the Maxwell relation for $p$ and $\mu$, which we define operationally 
for the system we study.  As we  explain below, the Maxwell relation 
provides an integrability condition 
for $p$ and $\mu$, and this yields a free energy function
extended to the NESS that we consider. The existence of this 
free energy function leads us to believe that there is 
an associated fluctuation relation. Indeed, our numerical experiments 
suggest that the compressibility can be expressed in terms of
density fluctuations even in certain non-equilibrium systems.

\paragraph*{Model:} 

Let $\sigma_i$ be an occupation variable defined on each site 
$i=(i_x,i_y)$ in a two-dimensional square lattice 
$\{(i_x,i_y)| 0 \le i_x \le L+1, 0 \le i_y \le M+1\}$.  
The variable $\sigma_i$ is 1 when the i-th site is occupied, and is 0 if it 
is empty. We assume  
periodic boundary conditions in the 
$x$-direction ( i.e. $\sigma_i=\sigma_j$ when 
when $i=j+(L,0)$), and 
no flux boundary conditions in the $y$-direction (i.e.
$\sigma_i=0$ when  $i_y=0, M+1$).
The array of all occupation variables 
$\{\sigma_i\}$ is denoted as $\sigma$ and called the ``configuration''.  

We study a driven lattice gas  with the Hamiltonian 
\begin{equation}
H( \sigma )
= - \sum_{ \bra i,j \ket} \sigma_i \sigma_j - E \sum_{i} i_x \sigma_i,
\end{equation}
where $\bra i,j \ket$ represents a nearest neighbor pair  
and $E$ is an external force \cite{KL}. 
The time evolution of $\sigma$ is described by the following
rule: At each time step, choose randomly a nearest neighbor pair 
$\bra i, j \ket$, and exchange the values of $\sigma_i$ and $\sigma_j$ 
with the  probability $c(i,j;\sigma)$  given by
\begin{equation}
c(i,j;\sigma)
=\frac{1}{1+\exp(\beta [H(\sigma^{ij})-H(\sigma)])},
\label{ex}
\end{equation}
where $\sigma^{ij}$ is the configuration obtained from $\sigma$ through
this exchange and $\beta$ is the inverse temperature. 
This exchange probability is called the heat 
bath method and is one of the most standard update rules satisfying the 
local detailed balance condition, which condition is regarded
to be natural in physical systems.
The particle number 
$N=\sum_{i}\sigma_i$ is  conserved throughout the time evolution. 
In this study, we fix $\beta=0.5$, in  which case the system is far from 
critical in the high temperature region.

\paragraph*{Pressure:}

\begin{figure}
\begin{center}
\includegraphics[width=8cm]{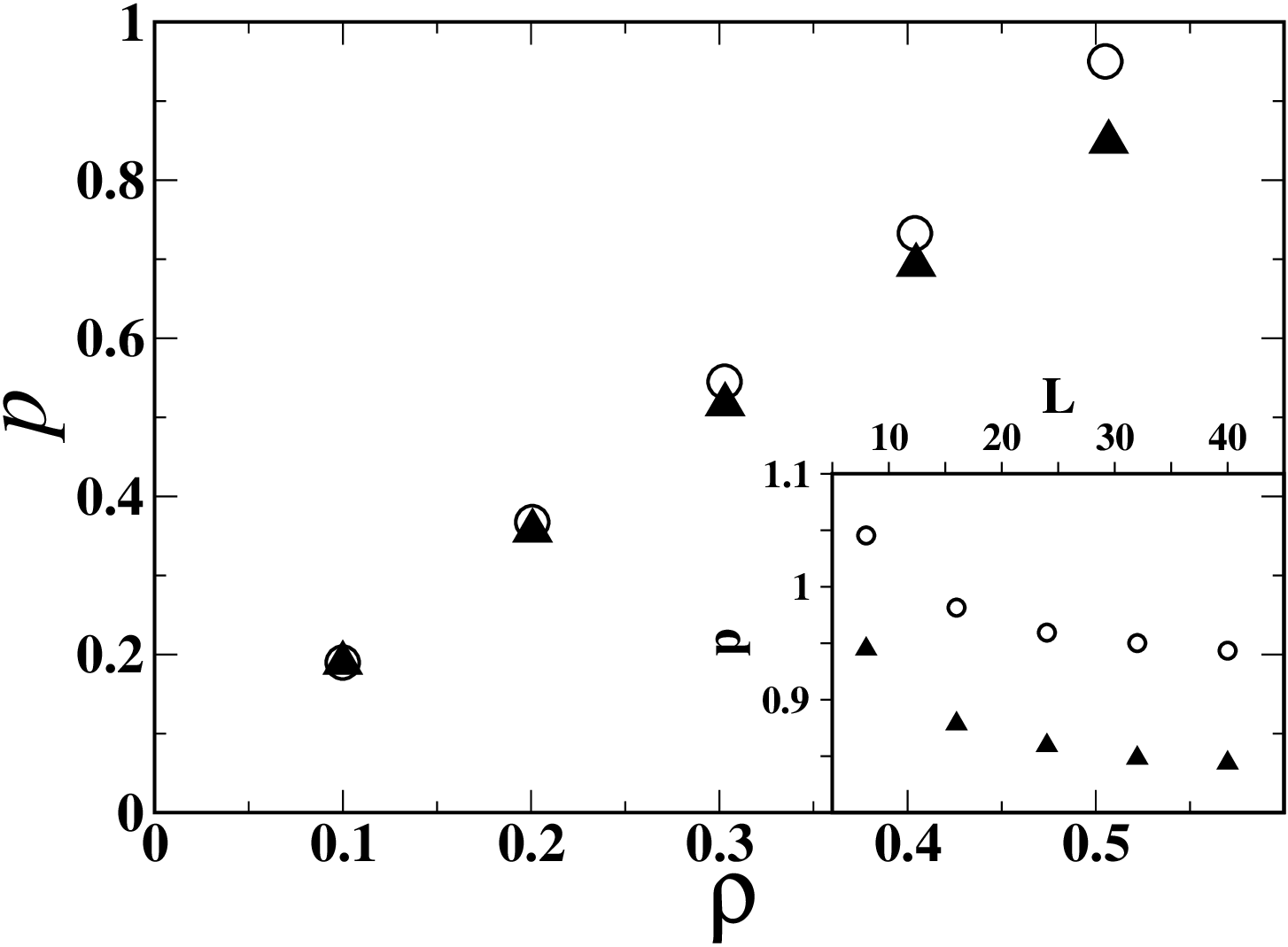}
\end{center}
\caption{Pressure as a function of the density averaged at the center 
of the system. The circular and triangular  symbols correspond to the 
cases $E=0$ and $E=10$, respectively. Here, $M=L=32$, $\Delta 
w=0.4 \times 10^{-5}$, $w_{\rm m}=30$
and $t_0=16000$  (msc).  The averaged values for 100 samples are plotted.
The statistical error bars are smaller than the symbols. 
The inset displays the $L$ dependence of the pressure with $N/L^2=0.5$ 
and $L/M=1$. The pressure seems to converge to a 
definite value in the thermodynamic limit.  
}
\label{p-rho}
\end{figure}

In the equilibrium case ($E=0$), the pressure is calculated by using
equilibrium statistical mechanics. However, because we do not know the
proper statistical measure in NESS, in this case we should define 
the pressure in an operational manner. 
Although the pressure is usually defined  as the normal force 
exerted on a unit area of a surface,  there is no quantity corresponding 
to the force in a lattice gas.  
We thus  define the pressure in terms of the quasi-static work 
required to change the system size. In order to allow for the
calculation of this quantity in our lattice gas  \cite{LG:pre}, we add 
a wall potential to the Hamiltonian $H(\sigma)$ in the form
\begin{equation}
H_w(\sigma)=H(\sigma)+\sum_{i: i_y=M} \sigma_i w .
\label{mh}
\end{equation}
Because of the boundary conditions we impose, no particles cannot exist
at sites with $i_y=M+1$, and therefore the system size in the $y$ 
direction is $M$. Now, according to (\ref{mh}), as $w$ is increased,
it becomes increasingly unlikely for particles to exist at sites with
$i_y=M$. When $w$ becomes sufficiently large, the average occupation
number for sites with $i_y=M$ can be considered zero. We denote the value of
$w$ beyond which this is the case as $w_{\rm m}$. Hence, in the process
that $w$ changes from 0 to $w_{\rm m}$, the effective system size in the
$y$ direction changes from $M$ to $M-1$. 
The quasi-static work performed to the system 
through this process is interpreted as the pressure 
multiplied by $L$. That is, the pressure $p$ is written as
\begin{equation}
p =\lim_{w_{\rm m} \to \infty}
\frac{1}{L} \int_0^{w_{\rm m}} dw  \sum_{\sigma} P_{N,M}^w(\sigma) 
\pder{H_w(\sigma)}{w},   
\label{pfor}
\end{equation}
where $P_{N,M}^w$ is the steady state distribution for a given value of $w$.
It is easily proved that this  definition of the pressure is equivalent
to statistical mechanical formula in equilibrium.

In the numerical experiments, values of $p$ are obtained in the following
way. Starting from random initial conditions, we carry out  the time evolution 
with $w=0$  for a sufficiently long time, say $t_0$. In this way, we 
obtain a steady state with $w=0$. Next, we increase the value of $w$ by 
a quantity $\Delta w$ per  Monte Carlo step per site (mcs) until $w$ reaches 
$w_{\rm m}$. Then, noting that $\partial{H_w(\sigma)}/\partial{w}= 
\sum_{i: i_y=M} \sigma_i$, we measure the value of 
$\sum_{i: i_y=M} \sigma_i$ as  $n(t)$ at  time $t$, 
where zero of the time is defined as the point at which 
$w$ starts to increase. 
In one process from $w=0$ to $w=w_{\rm m}$, we calculate
\begin{equation}
p =\frac{\Delta w}{L}  \sum_{t=1}^{t_{\rm m}} n(t),
\end{equation}
where $t_{\rm m}$ is the time at which $w=w_{\rm m}$. 
Then, determining  carefully how the statistical distribution of $p$  depends 
on $w_{\rm m}$ and $\Delta w$, we estimate the value of $p$ in the 
limit $w_{\rm m} \rightarrow \infty$,  $\Delta w \rightarrow 0$
\cite{Ha-S}. 

In Fig. \ref{p-rho}, we display  an example of measured values of 
the pressure for densities  in systems with $E=0$ and $E=10$. 
It is important to note that our analysis is not restricted to  systems 
near equilibrium. Indeed, the system with $E=10$ is close to 
the strong field limit, in which the particle current is 
saturated to a constant value, and the equation of state for $E=10$
clearly deviates from the equilibrium one. 
This difference shows that the statistical distribution 
in the $y$ direction differs from the equilibrium one. 
Also,
the pressure becomes an intensive variable in the thermodynamic limit,
as seen in the inset of Fig. \ref{p-rho}.

\paragraph*{Chemical potential:}

The chemical potential is measured by placing a particle reservoir 
in contact with the system in  the  direction transversal to the 
driving field.  We first assume that the chemical potential of the reservoir, 
$\mu_{\rm R}(T,\rho_{\rm R})$, is known.  We also assume that 
there exists a chemical potential $\mu$ in this NESS that is a function of 
$T$, $\rho$ and $E$. {}From the definition of the chemical potential,
we should have $\mu(T,\rho,E)=\mu_{\rm R}(T,\rho_{\rm R})$. Then,  using 
$\mu(T,1-\rho,E)=\mu_{\rm R}(T,1-\rho_{\rm R})$,  which holds due to 
the particle-hole symmetry, the equality $\mu(T,\rho=1/2, E)
=\mu_{\rm R}(T,\rho_{\rm R}=1/2)$ is derived.  Thus, by measuring
$\partial \mu(T,\rho, E)/\partial \rho$ for all values of $\rho$
without contacting a particle reservoir, we 
can determine the form of the function  $\mu(T,\rho, E)$.

In order to measure $\partial \mu(T,\rho, E)/\partial \rho$ 
numerically, we add the one-body potential term $\sum_i \sigma_i 
\phi_i$ to the Hamiltonian $H(\sigma)$, where  $\phi_i = \Delta \phi$ 
for $i_y \ge M/2+1$, and $\phi_i = 0$ for $ i_y \le M/2$.
We then measure the density profile along the $y$ direction, in which 
there are two flat regions, $1 \ll i_y  \ll M/2$ and 
$M/2+1 \ll i_y  \ll M$. We denote the density in  the region 
$1 \ll i_y  \ll M/2$ 
as $\rho_1$ and the density in the region 
$M/2+1 \ll i_y  \ll M$ as $\rho_2$.  When 
$\Delta \phi$ is sufficiently small, the chemical potentials of the two regions
are equal by the  definition of the chemical potential. By taking into
account the shift in the potential energy, this condition can be written as
\begin{equation}
\mu(T,\rho_1, E) = \mu(T,\rho_2, E)+\Delta \phi.
\label{cpot:0}
\end{equation}
We thus obtain 
\begin{equation}
\left.   \pder{\mu(T,\rho, E)}{\rho}\right|_{\rho=\bar \rho} =
\lim_{\Delta \phi \to 0}\frac{\Delta \phi}{\Delta \rho},
\label{cpot:1}
\end{equation}
where $\Delta \rho =\rho_1-\rho_2$ and $\bar \rho=(\rho_1+\rho_2)/2.$
Measuring $\Delta \rho$ and $\bar \rho$ for several values of 
$\Delta \phi$, we can evaluate the right-hand side of 
(\ref{cpot:1}) \cite{Ha-S}.

\paragraph*{Free energy:}

\begin{figure}
\begin{center}
\includegraphics[width=6cm]{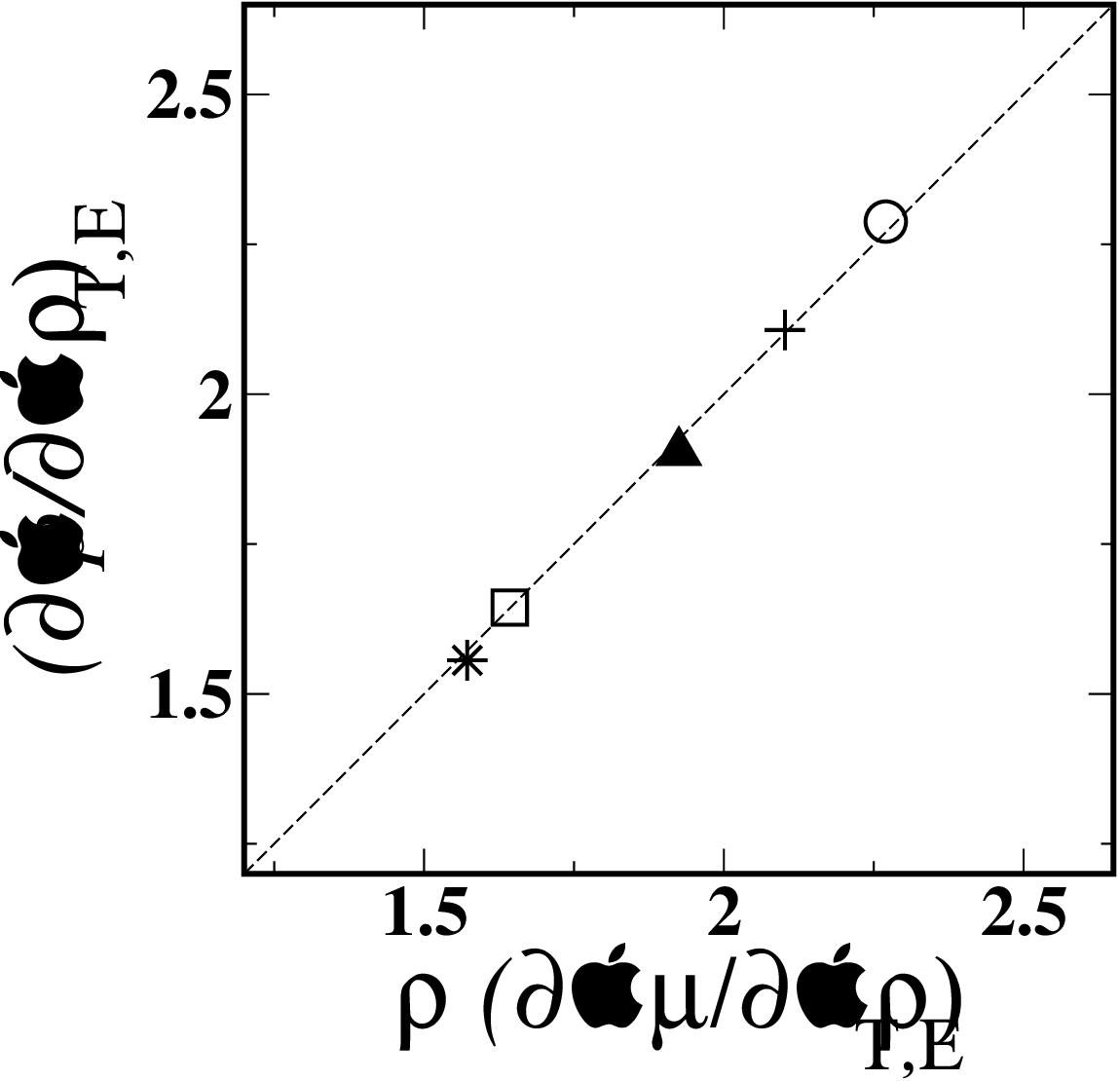}
\end{center}
\caption{
$\partial p(T,\rho, E)/\partial \rho$ versus 
$\rho \partial \mu(T,\rho, E)/\partial \rho$
for several values of $\rho$ and $E$. 
The triangle, square, star, plus, and circle corresponds 
to $(\rho,E)$=$(0.5 ,10)$, $(0.4,10)$, $(0.3,10)$, $(0.5, 3.0)$ and 
$(0.5,0.0)$, respectively. Here, 
$L=32$, and $M=32$ for $\partial p(T,\rho, E)/\partial \rho$ and 
$M=64$ for $\rho \partial \mu(T,\rho, E)/\partial \rho$.
The statistical error bars are smaller than the symbols.
The dotted line corresponds to the Maxwell relation.
}
\label{fig:max}
\end{figure}

We begin by conjecturing that the relation
\begin{equation}
\pder{p(T,\rho,E)}{\rho}=\rho\pder{\mu (T,\rho,E)}{\rho},
\label{max}
\end{equation}
which holds in equilibrium, holds also in our NESS. This 
is equivalent to the Maxwell relation, because  $p$ and $\mu$ are
numerically confirmed to be intensive. Here, we estimate
the value of $\partial p/\partial\rho$ (at $\rho=0.5$, for example) 
by calculating 
values of $p$  for several values of $\rho$ in a small 
interval around $\rho=0.5$ \cite{Ha-S}. The results  summarized in 
Fig. \ref{fig:max}  suggest the validity of the equality 
(\ref{max}). If indeed this relation does hold, its implication is
significant, because when we define the quantity $F(T,M,N, E)$ as 
\begin{equation}
F(T,M,N,E)=N \mu \left(T,\frac{N}{ML},E\right)
         - ML p\left(T,\frac{N}{ML},E\right) ,
\label{free}
\end{equation}
$p$ and $\mu$ become 
\begin{eqnarray}
p &=&  - \frac{1}{L} \pder{F(T,M,N,E)}{M},
\label{pf} \\
\mu &=&  \pder{F(T,M,N,E)}{N}.
\label{muf}
\end{eqnarray}
That is, $F(T,M,N,E)$ can be regarded as the free energy
extended to the present NESS. 
Using the single function $F(T,M,N,E)$, we can derive
various thermodynamic relations, including the Clapeyron law, under 
non-equilibrium conditions.

\paragraph*{Fluctuation relation:}

\begin{figure}
\begin{center}
\includegraphics[width=8cm]{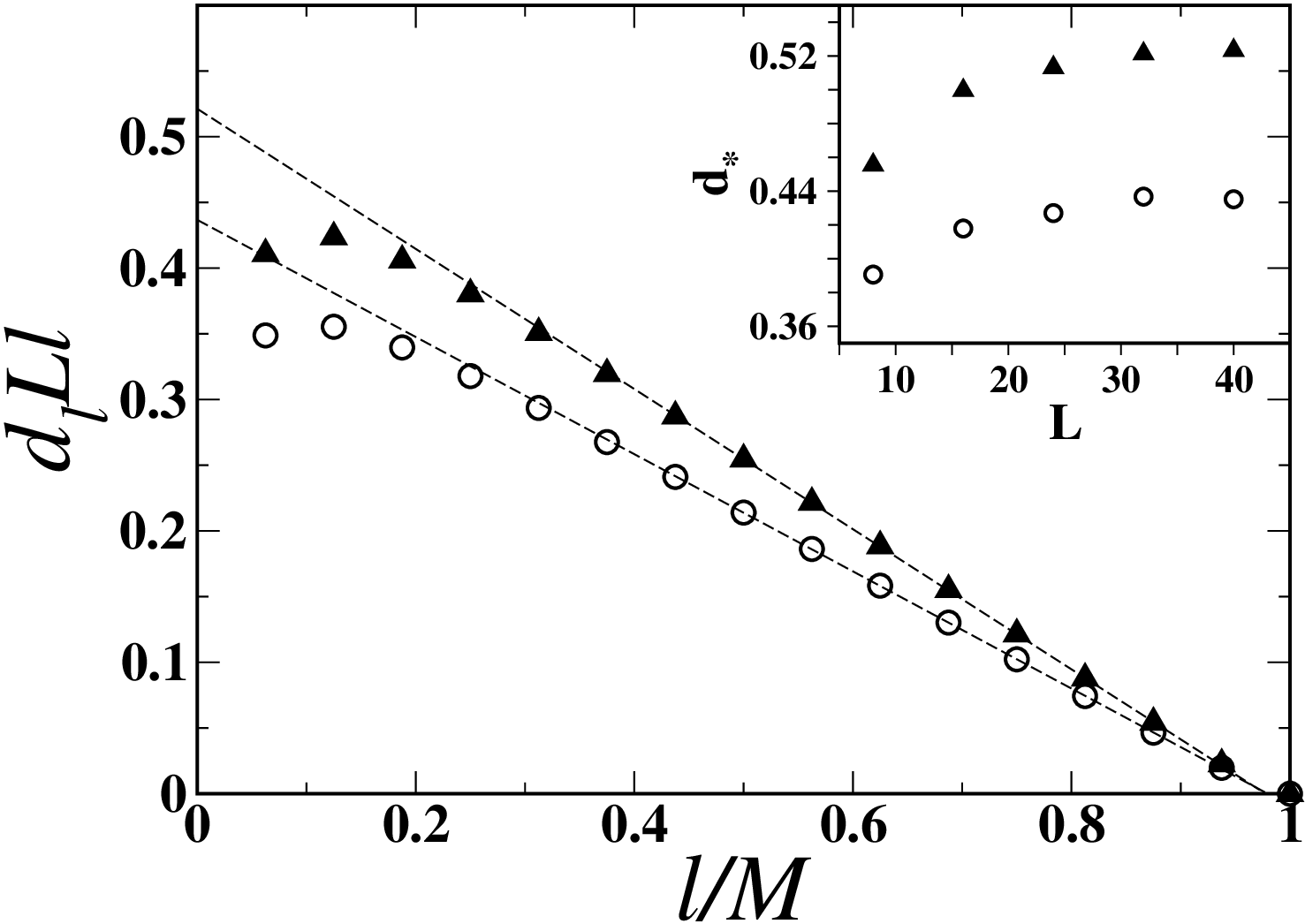}
\end{center}
\caption{$d_\ell L \ell$ as a function of  $\ell/M$. Here, $L=M=32$. 
The statistical error bars are smaller than the symbols. The dotted 
line is obtained as the best fit line of the form $a \ell/M  +b$ 
in the region $\ell/M \in [0.4,0.8]$, and $d_*$ is evaluated as 
the value of $b$.  The inset displays the $L$ dependence of $d_*$ with 
the values $N/L^2=0.5$ and $L/M=1$ fixed. 
$d_*$ seems to converge to a definite value in the thermodynamic limit.  
}
\label{d-l}
\end{figure}

Consider the strip 
$
\Omega_l =\{(i_x, i_y)|1 \le i_x \le L, M/2-\ell/2-1 \le i_y \le M/2+\ell/2 \},
$
and define the density variable 
$
\rho_\ell= \sum_{i \in \Omega_l } \sigma_i/{|\Omega_l|} 
$, which is coarse-grained 
over the strip $\Omega_l$. Let $\xi$ be the correlation length of density
fluctuations in the y-direction. 
We define the free energy density $f(\rho)$ for  fixed $(T,E)$
by $F(T,M,N,E) = M f(N/M)$ in the thermodynamic limit. 
It is then conjectured that the probability distribution of the density 
$\rho_\ell$ with $\xi \ll \ell \ll M $ can be written as
\begin{equation}
P(\rho_\ell=\rho)\simeq \exp(-\beta \ell [f(\rho)-f(\bar \rho)]),
\label{LD}
\end{equation}
where $\bar \rho$ is the thermodynamic density. In the equilibrium case, 
such a form can be derived from a fundamental principle of statistical 
mechanics. 
Although there is no general proof of this form in the case of NESS, 
(\ref{LD}) seems plausible in the present case, because we have been
able to define a free energy \cite{eyink}. 

Recently, for some non-equilibrium lattice gases, the large deviation 
functionals of density fluctuations have been  rigorously derived in non-local 
forms \cite{DLS}. These nonlocal forms are related to long-range 
correlations that exist generically in NESS \cite{LRC}. 
Here we shall not discuss this important issue further, and simply state 
our numerical finding that the scaling form (\ref{LD}) 
has been clearly observed 
provided that one examines the density fluctuation in the strip 
$\Omega_\ell$.

If (\ref{LD}) is valid (at least locally),  the fluctuation relation 
\begin{equation}
\beta \pder{\mu}{\rho}=\frac{1}{L\ell \bra (\rho-\bar \rho)^2\ket}
\label{fr}
\end{equation}
can be derived. This relation is known to be valid for describing fluctuations
about equilibrium states, but it is not known whether it is valid in NESS.
To study this point, we investigate  (\ref{fr})  numerically.

\begin{figure}
\begin{center}
\includegraphics[width=6cm]{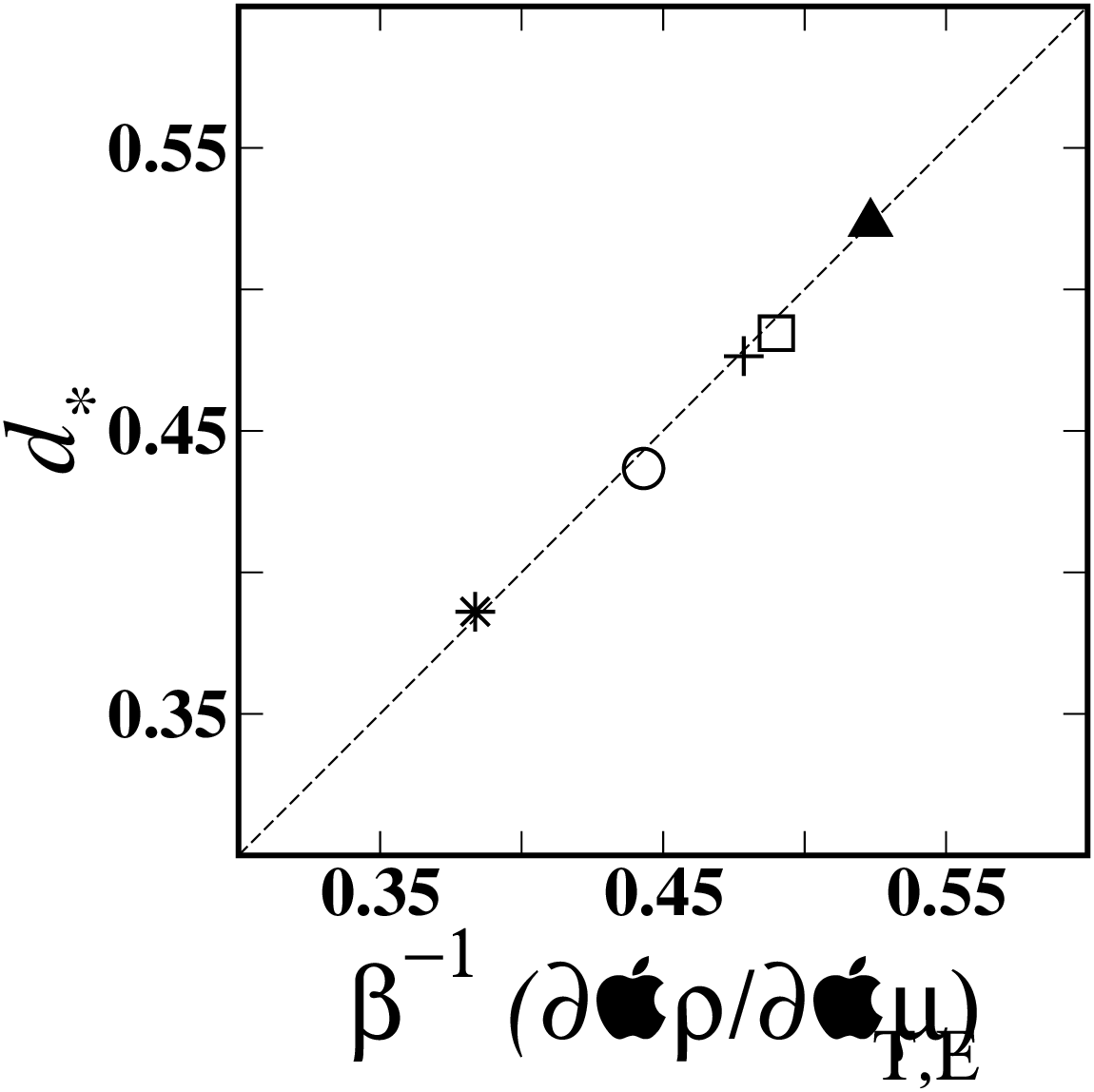}
\end{center}
\caption{
$d_*$ versus  $\beta^{-1} \partial \rho/\partial\mu$ 
for several values of $(\rho,E)$. The triangle, square, star, 
plus, and circle correspond to $(\rho,E)=(0.5 ,10)$, $(0.4,10)$, 
$(0.3,10)$, $(0.5, 3.0)$ and $(0.5,0.0)$, respectively. 
Here, $L=32$, and  $M=32$ for $d_*$ and $M=64$ for 
$\beta^{-1} \partial \rho/\partial \mu$.
The statistical error bars are smaller than the symbols.
The dotted line corresponds to the fluctuation relation.
}
\label{f-r}
\end{figure}

We first note the asymptotic form 
\begin{equation}
d_\ell\equiv
\bra \rho_l^2 \ket- \bra \rho_l \ket^2\simeq 
d_* \frac{1}{L \ell} \frac{M-\ell}{M} 
\label{scaling2}
\end{equation}
for $\xi \ll \ell \ll  M$ (see Fig. \ref{d-l}), where 
$d_*$ is defined.
According to this, 
$L\ell \bra (\rho-\bar \rho)^2\ket$ in (\ref{fr}) should
correspond to $d_*$.  The values  of $d_*$ and 
$\beta^{-1} \partial \rho/\partial\mu$  are plotted 
in Fig . \ref{f-r}. This result suggests the 
validity of (\ref{fr}). In addition, combining (\ref{fr}) with (\ref{max}), 
the relation we obtain between the compressibility and density fluctuations
is the same as that existing in equilibrium.

We remark here that the asymptotic form (\ref{scaling2}) can 
be understood 
by considering the following simple situation.  
Consider $n$ random variables $x_i$ ($1\le i \le n$), with the conservation 
constraint $\sum_{i=1}^n x_i=0$.
The statistical properties of $x_i$ are given by $E(x_i)=0$ and 
$E(x_ix_j)=\delta_{ij}$, where $E(x)$ represents the expected value of 
the random variable $x$. Let $X_k$ be the  partial sum of $k$ elements 
randomly chosen from the set $\{x_i \}$. Then, the probability 
of $X_k$ with large $k$ and large $n-k$ is given by the Gaussian 
distribution with $E(X_k)=0$ and $E(X_k^2)=k (n-k)/n$ \cite{sol}. 
This supports the form of (\ref{scaling2}).

\paragraph*{Discussion:}

We have presented new relations obtained using numerical experiments on 
a driven lattice gas. 
We hope that the present 
study can be extended to a wider variety of systems. 
Also, finding a connection between the entropy, which is defined 
as $-(\partial F/\partial T)_{M,N,E}$ 
in our formulation, and  the Shannon entropy would be important 
in  future construction of a  statistical mechanical theory of NESS \cite{HS}. 

The authors acknowledge H. Tasaki for stimulating discussions on steady 
state thermodynamics and driven lattice gases.  They also thank Y. Oono
and A. Shimizu for fruitful discussions on NESS. This work was supported 
by grants from the Ministry of Education, 
Science, Sports and Culture of Japan, No. 14654064.


\begin{thebibliography}{10}
\bibitem{LG:pre}

We consider the pressure only in the  direction 
transversal to the driving 
field. In general, the pressure tensor becomes anisotropic in NESS. 
However, we do not know how to measure the pressure in the 
direction parallel to the driving field.
\bibitem{sol}

If there is no conservation constraint, the combined probability 
of $X_k$ and $X_{n-k}$, $P(X_k, X_{n-k})$, approaches  a product 
of Gaussian distributions as $k$ and $n-k$ become large. The asymptotic 
distribution of $X_k$ with the conservation constraint is simply given by
$P(X_k, -X_{k})$.  
\end{thebibliography}
\end{document}